\documentclass[12pt,a4paper]{article}
\usepackage[utf8]{inputenc}
\usepackage[T1]{fontenc}
\usepackage{newtxtext,newtxmath}
\usepackage[margin=1in]{geometry}
\usepackage{amsmath,amsfonts}
\usepackage{graphicx}
\usepackage{float}
\usepackage{placeins}
\usepackage{subcaption}
\usepackage{setspace}
\usepackage{caption}
\usepackage{color}
\newcommand{\rev}[1]{{\color{black}#1}}

\newlength{\figurepanelheight}
\begin{document}

\begin{center}
{\LARGE \bfseries  Topological Magnons and Altermagnetism in Stacked van der Waals Bilayer CrI$_3$}

\vspace{1.2em}
{\large Lu Xiao$^{1}$, Meng-Han Zhang$^{1}$, Dao-Xin Yao$^{1,\dagger}$}

\vspace{0.6em}
{\normalsize $^{1}$Guangdong Provincial Key Laboratory of Magnetoelectric Physics and Devices, State Key Laboratory of Optoelectronic Materials and Technologies, Institute of Neutron Science
and Technology, School of Physics, Sun Yat-Sen University, Guangzhou, 510275, China \\
$^{1}$These authors contributed equally to this work.\\
$^{\dagger}$Corresponding author: yaodaox@mail.sysu.edu.cn}
\end{center}

\vspace{1em}

\begin{abstract}
Stacked van der Waals magnets provide a tunable route to altermagnetism, a compensated magnetic order characterized by momentum-dependent spin splitting, and to the topological magnetic excitations that such order can host. In bilayer CrI$_3$, first-principles calculations and linear spin-wave theory reveal stacking-controlled altermagnetic order and associated magnon band topology. Combining band-representation analysis with calculations of the dynamic structure factor relevant to inelastic neutron scattering, we further characterize the chirally split topological magnons and quantify their energy corrections and lifetimes using a many-body Green's-function approach. The interlayer magnetic ground state is highly sensitive to the stacking geometry, thereby controlling the magnon band topology and transport responses. We further show that magnon--magnon interactions renormalize the magnon dispersion and dynamic structure factor, with a particular focus on magnon decay. Using van der Waals bilayer CrI$_3$ as a representative platform, our results establish stacking engineering as a structural route for tuning altermagnetism and associated topological magnon excitations, opening avenues toward stacking-controlled spintronic and magnonic devices.
\end{abstract}

\vspace{1em}
\section*{Introduction}

Stacking in bilayer van der Waals (vdW) magnets is a symmetry operation on the composite crystal rather than a geometric detail, and it is the control variable exploited here. The registry between the layers fixes which lattice operations survive the stacking\cite{chen2024}, and therefore whether the interlayer exchange is ferromagnetic or antiferromagnetic\cite{jang2019,soriano2019} and, with it, whether the magnetic ground state is a conventional collinear antiferromagnet, in which the nonrelativistic spin splitting vanishes identically, or an altermagnet, in which the staggered order produces chirality-split magnon bands and Weyl magnons\cite{cui2026twisted,mook2016,su2017,Guo_2025,lin2026chiral,zhang2026collinear}. In stacked chromium triiodide (CrI$_3$), the interplay between the Dzyaloshinskii--Moriya interaction (DMI), magnetic anisotropy and the interlayer coupling gives rise to gapped magnon bands with nontrivial Chern numbers, so that chiral edge magnons and a finite thermal Hall response emerge\cite{mcclarty2022,mook2014,katsura2010,shindou2013,chen2018}. Because the registry can be varied continuously by twist angle, strain, electrostatic gating, intercalation and layer number without altering the chemical composition, one and the same bilayer is tunable across this phase diagram; where the registry varies periodically, the moir\'e texture\cite{ghader2022,kim2022} folds and hybridizes the acoustic and optical branches into minibands, displaces replica nodes to $\pm\mathbf{q}_M$, and supplies the additional momentum scale through which magnons scatter. These predictions are falsifiable by inelastic neutron spectroscopy, Raman scattering and domain-resolved imaging.

Chromium triiodide (CrI$_3$) is an experimentally realized van der Waals (vdW) honeycomb ferromagnet and has become a paradigmatic platform for topological magnonics\cite{huang2017,pershoguba2018,chen2018,chen2021}. In its staggered bilayer, altermagnetic order lifts the effective time-reversal constraint on Berry curvature and generates a nonrelativistic, momentum-dependent spin splitting\cite{cui2026twisted}, and inelastic neutron scattering (INS) resolves the substantial lifetime broadening of the associated magnon modes\cite{chen2018,schneeloch2024}. Self-consistent renormalized (SCR) spin-wave theory captures the overall renormalization of the CrI$_3$ magnon spectrum\cite{mkhitaryan2021}, but the microscopic vertex that governs this renormalization, and through it the topology and the lifetime of these topological excitations, remains unquantified. Whether the renormalized excitations remain identifiable as quasiparticles is decided by the quartic terms of the Hamiltonian\cite{zhitomirsky2013,sourounis2024}. Thermal fluctuations populate higher-energy magnon states, so the self-energy corrections acquire pronounced momentum and temperature dependence\cite{dyson1956a,dyson1956b,zhitomirsky2013,pershoguba2018,nikitin2022,garciagaitan2025}, and at finite temperature these interactions manifest themselves in the decay of magnon lifetimes, the renormalization of bandwidths\cite{zhitomirsky2013,chen2023,sourounis2024} and, potentially, in sign reversals of the transverse spin Nernst conductivity\cite{cheng2016}.

Isolating the vertex weight $W(\mathbf{p},\mathbf{q};\mathbf{k})$ shows that these corrections divide the spectrum into two regions of qualitatively different behaviour. Around the topological gap, the first- and second-order self-energies are small and smooth in momentum, so the gap, the branch-resolved Chern numbers and the bulk--edge correspondence survive renormalization, because a smooth, momentum-local correction cannot close a gap. Near the Weyl points and the Brillouin-zone boundary, by contrast, the vertex becomes large and momentum selective, and the pronounced dips it generates are reproduced only when the second-order self-energy is included. The chirality-dependent part of the spectrum retains its sixth-order angular harmonic, so the $i$-wave form factor survives with renormalized amplitude and phase, and the chiral splitting is reduced but not destroyed with respect to linear spin-wave theory. The nonrelativistic magnon splitting is thus qualitatively robust and functionally usable, even though its quantitative magnitude departs from the bare spin-wave value. Because the Weyl dip flattens the local slope of the dispersion, and hence the group velocity, magnon transport is localized in the immediate vicinity of the nodes. The spin Nernst conductivity\cite{cheng2016,zyuzin2016}, which weights this group velocity by both the Berry curvature and the thermal occupation, therefore reports a thermal reconstruction of the dispersion rather than a topological transition: the sign reversals found here upon the bilayer reflect a redistribution of Berry-curvature-weighted spectral weight.

\section*{Results}

\subsection*{Spin-wave model and Altermagnetism}

\begin{figure}[!t]
\centering
\begin{subfigure}[t]{0.48\textwidth}
\centering
\includegraphics[width=\linewidth]{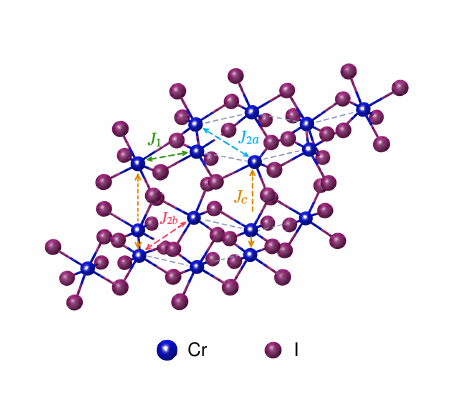}
\caption{}
\label{fig1a}
\end{subfigure}
\hfill
\begin{subfigure}[t]{0.48\textwidth}
\centering
\includegraphics[width=\linewidth]{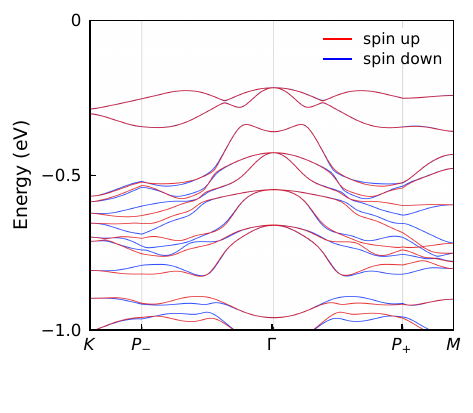}
\caption{}
\label{fig1b}
\end{subfigure}
\par\medskip
\begin{subfigure}[t]{0.48\textwidth}
\centering
\includegraphics[width=\linewidth]{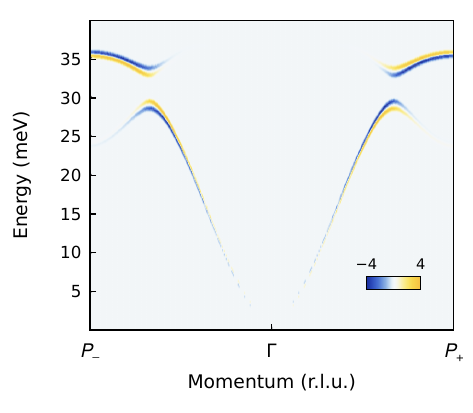}
\caption{}
\label{fig1c}
\end{subfigure}
\hfill
\begin{subfigure}[t]{0.48\textwidth}
\centering
\includegraphics[width=\linewidth]{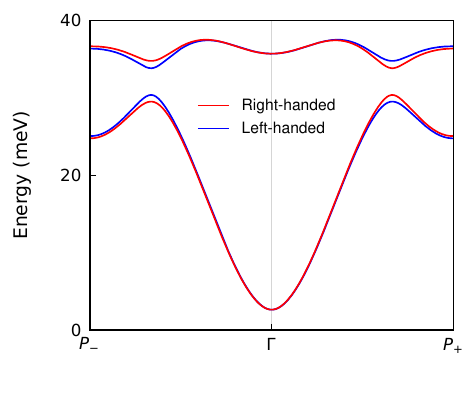}
\caption{}
\label{fig1d}
\end{subfigure}
\caption{\textbf{Electronic spin splitting and magnon chirality in bilayer CrI$_3$.} 
(a) Crystal structure and exchange paths of the bilayer spin model. 
(b) Electronic spin splitting without spin--orbit coupling(SOC). Red and blue denote spin up and spin down. 
(c) Dynamical spin chirality, where the colorbar denotes $\mathrm{Im}[M^{xy}-M^{yx}]$, resolving the circular polarization and chirality of the magnon excitations.
(d) Momentum-dependent splitting of right-handed (red) and left-handed (blue) magnon branches.}
\label{fig1}
\end{figure}

Among layered van der Waals magnets, bilayer CrI$_3$ provides a particularly suitable platform for realizing topological Weyl magnons because its interlayer magnetic order can be switched between ferromagnetic and antiferromagnetic configurations\cite{zhang2026edge}. 
In the staggered magnetic bilayer, the Cr moments in the two layers are aligned antiparallel, and the Cr$^{3+}$ ions possess three unpaired $d$ electrons in an approximately octahedral coordination environment. Each Cr$^{3+}$ ion carries a localized spin moment on the honeycomb lattice, 
and the magnetic excitations are therefore induced from the local spin representation at the Cr Wyckoff positions $2c$ and $2d$, where the site symmetry $^{1}3..^{\infty m}1$ is generated by a threefold rotation $C_{3z}$ accompanied by the identity operation in spin space\cite{chen2024,yu2026findspingroup}.

The symmetry analysis reveals that the magnetic order retains a continuous internal spin symmetry $O(2)=SO(2)\rtimes Z_2$, allowing a continuous spin rotation $U_z(\phi)$ about the spin $z$ axis\cite{liu2022spingroup,chen2024,liu2026ossz,yu2026findspingroup}. 
The semidirect-product structure follows from the fact that the mirror operation reverses the sense of the continuous spin rotation $\rev{mU_z(\phi)m^{-1}=U_z(-\phi)}$. Despite the absence of a net moment, 
the staggered interlayer order fundamentally modifies the magnetic symmetry. The antiferromagnetically stacked bilayer breaks the time-reversal symmetry that relates the two magnon branches\cite{smejkal2022a,smejkal2022b,eto2025}, 
while retaining the lattice symmetries compatible with the staggered spin configuration. 
As a result, the altermagnetic structure is reduced to the spin space group $P^{1}3^{-1}2^{1}1^{\infty\mathrm{m}_{1}}$ (143.150)\cite{liu2022spingroup,chen2024,liu2026ossz,yu2026findspingroup}, where the magnon band representation from the relevant Wyckoff position decomposes into the following symmetry-adapted irreducible representations:


\begin{equation}
\begin{aligned}
\left(^{1}E\right)_{2c}\uparrow P321&=\Gamma_{1}^{+}(2)\oplus K_{1}(2)\oplus M_{1}^{+}(2),\\
\left(^{1}E\right)_{2d}\uparrow P321
&=\Gamma_{1}^{+}(2)\oplus K_{2}(2)\oplus M_{1}^{+}(2).
\end{aligned}
\end{equation}

The magnon band representation induced from the Cr spin degrees of freedom decomposes into symmetry-resolved irreducible representations at the high-symmetry momenta of the Brillouin zone. 
These irreducible representations dictate symmetry-allowed band inversions between magnon branches, providing isolated linear crossings as Weyl magnons\cite{mook2016,su2017,gohlke2023}. 
Rhombohedral stacking favors ferromagnetic interlayer coupling, whereas monoclinic stacking can support antiferromagnetic interlayer order\cite{jang2019,soriano2019}.
The magnetic properties of bilayer CrI$_3$ are described by the anisotropic Heisenberg model on the honeycomb lattice (Fig.~\ref{fig1})\cite{chen2018,lado2017,soriano2019},

\begin{equation}
\begin{aligned}
\mathcal{H}={}&\rev{-J_{1}}\sum_{\langle mn\rangle} \mathbf{S}_{m}\cdot\mathbf{S}_{n}
\rev{-}\sum_{\nu=a,b}J_{2\nu}\sum_{\langle\langle mn\rangle\rangle_{\nu}} \mathbf{S}_{m}\cdot\mathbf{S}_{n}
\rev{-J_{3}}\sum_{\langle\langle\langle mn\rangle\rangle\rangle} \mathbf{S}_{m}\cdot\mathbf{S}_{n}\\
&+J_{c}\sum_{\langle mn\rangle} \mathbf{S}_{m}\cdot\mathbf{S}_{n}
+\sum_{\langle mn\rangle} \rev{D_z}\,\hat{z}\cdot(\mathbf{S}_{m}\times\mathbf{S}_{n})
-K\sum_{\langle mn\rangle} S_{m}^{z}S_{n}^{z}+h_{B}\sum_{m} S_{m}^{z},
\end{aligned}
\end{equation}

where $J_1$ and $\rev{J_3}$ are the first- and third-nearest-neighbour exchange couplings, respectively, while $J_{2a}$ and $J_{2b}$ denote two different values of the in-plane second-nearest-neighbour exchange coupling $J_2$. \rev{The parameter $J_c$ denotes the interlayer exchange coupling.} Here, $\langle\langle mn\rangle\rangle_{\nu}$ denotes the corresponding in-plane second-neighbour bond set, with $\nu=a,b$. Furthermore, $\rev{D_z}$ is the out-of-plane DMI, $K$ is the easy-axis anisotropy and $h_B$ is the Zeeman field.
The ground state is a fully polarized ferromagnet, whose robustness follows from the absence of frustration in the mean-field description. 
To study the magnetic excitations, we map the spin Hamiltonian onto a bosonic problem using the higher-order Holstein--Primakoff (HP) transformation, where the higher-order terms are retained to account for magnon--magnon interactions beyond linear spin-wave theory.\cite{dyson1956a,zhitomirsky2013,yao2008,yao2010}

\begin{equation}
\begin{aligned}
S_{m}^{+}=\sqrt{2S}\left[ a_{m}-\frac{1}{2}\frac{a_{m}^{\dagger}a_{m}a_{m}}{2S}\right],\quad
S_{m}^{-}=\sqrt{2S}\left[ a_{m}^{\dagger}-\frac{1}{2}\frac{a_{m}^{\dagger}a_{m}^{\dagger}a_{m}}{2S}\right],\quad
S_{m}^{z}=S-a_{m}^{\dagger}a_{m},\\
S_{n}^{+}=\sqrt{2S}\left[ b_{n}^{\dagger}-\frac{1}{2}\frac{b_{n}^{\dagger}b_{n}^{\dagger}b_{n}}{2S}\right],\quad
S_{n}^{-}=\sqrt{2S}\left[ b_{n}-\frac{1}{2}\frac{b_{n}^{\dagger}b_{n}b_{n}}{2S}\right],\quad
S_{n}^{z}=-S+b_{n}^{\dagger}b_{n},\\
\end{aligned}
\end{equation}

where $a_{m}^{\dagger}$ ($a_m$) creates (annihilates) a magnon on sublattice $a$, whereas $b_{n}^{\dagger}$ ($b_n$) creates (annihilates) a magnon on sublattice $b$. 
We expand the exchange and DMI terms to quartic order in the boson operators, performing the Fourier transformation for the quadratic part of the Hamiltonian,

\begin{equation}
\begin{split}
H(\mathbf{k})
={}&
\frac{d_{1a}+d_{1b}+d_{2a}+d_{2b}}{4}\,\tau_0\otimes\sigma_0
+\frac{d_{1a}-d_{1b}+d_{2a}-d_{2b}}{4}\,\tau_0\otimes\sigma_z
\\
&+\frac{d_{1a}+d_{1b}-d_{2a}-d_{2b}}{4}\,\tau_z\otimes\sigma_0
+\frac{d_{1a}-d_{1b}-d_{2a}+d_{2b}}{4}\,\tau_z\otimes\sigma_z
\\
&+\frac{J_c}{2}\left(\tau_x\otimes\sigma_x-\tau_y\otimes\sigma_y\right)
-\frac{M_1+M_2}{2}\,\tau_0\otimes\sigma_x
\\
&+\frac{N_1+N_2}{2}\,\tau_0\otimes\sigma_y
-\frac{M_1-M_2}{2}\,\tau_z\otimes\sigma_x
+\frac{N_1-N_2}{2}\,\tau_z\otimes\sigma_y,
\end{split}
\end{equation}

where $M_1+\mathrm{i}N_1=e^{-\mathrm{i}k_y}+2e^{\mathrm{i}k_y/2}\cos(\sqrt{3}k_x/2)$, $M_2+\mathrm{i}N_2=e^{\mathrm{i}k_x}+2e^{-\mathrm{i}k_x/2}\cos(\sqrt{3}k_y/2)$.
The Hamiltonian coefficients are defined by $d_{1a,1b}=3J_1+\tfrac{J_{2a,2b}}{2}(15-M_1^2-N_1^2)$, 
and $d_{2a,2b}=3J_1+\tfrac{J_{2a,2b}}{2}(15-M_2^2-N_2^2)$.
The magnon energies are then obtained from the eigenvalues of the non-Hermitian dynamical matrix $\Xi H(\mathbf{k})$ with $\Xi = \tau_z \otimes \sigma_0$, which accounts for the bosonic commutation relations in the Bogoliubov–de Gennes (BdG) formalism. 
We perform first-principles calculations for bilayer CrI$_3$ with intralayer ferromagnetic and interlayer antiferromagnetic ordering. As shown in Fig.~\ref{fig1b}, the resulting electronic band structure exhibits momentum-dependent spin splitting even in the absence of SOC, providing a characteristic electronic signature of the altermagnetic state\cite{zhang2024dirac}. The calculated exchange coupling parameters and single-ion anisotropy are summarized in Table~\ref{tab:exchange}.
\begin{table}[htbp]
\centering
\caption{\textbf{Microscopic parameters of the bilayer CrI$_3$ spin model.} Exchange couplings and single-ion anisotropy are given in meV; the corresponding bond lengths are in \AA.}
\label{tab:exchange}
\setlength{\tabcolsep}{4pt}
\begin{tabular}{c c c c c c c c c}
\hline\hline
Bilayer CrI$_3$
& $J_1$
& $J_{2a}$
& $J_{2b}$
& $J_{3a}$
& $J_{3b}$
& $J_{3c}$
& $J_c$
& $K_z$ \\
\hline
Bond length (\AA)
& 4.248
& 7.358
& 7.358
& 8.205
& 8.213
& 8.204
& 7.010
& \textemdash{} \\
Value (meV)
& $-2.146$
& $-0.568$
& $-0.546$
& $0.072$
& $0.055$
& $0.072$
& $1.849$
& $-0.258$ \\
\hline\hline
\end{tabular}
\end{table}

\subsection*{Magnon Band Topology}

\begin{figure}[H]
\centering
\begin{subfigure}[t]{0.48\textwidth}
\centering
\includegraphics[width=\linewidth]{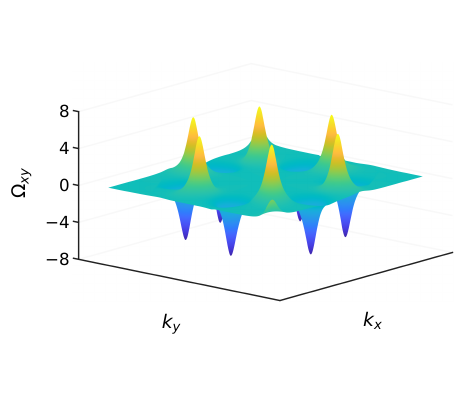}
\caption{}\label{fig:bands-b}
\end{subfigure}
\hfill
\begin{subfigure}[t]{0.48\textwidth}
\centering
\includegraphics[width=\linewidth]{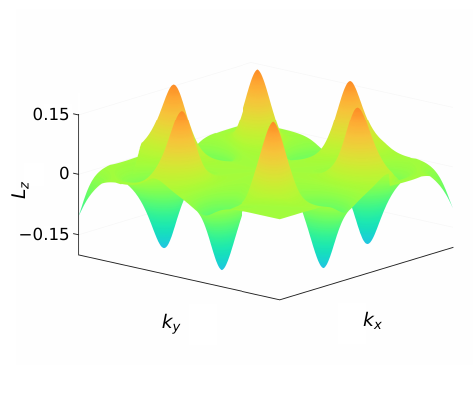}
\caption{}\label{fig:bands-c}
\end{subfigure}
\caption{\textbf{Geometric and orbital character of magnons.} (a) Momentum-resolved Berry curvature $\Omega_{xy}$, which governs the intrinsic transverse magnon response. (b) Orbital angular momentum $L_z$, with opposite signs in different momentum regions.}
\label{fig2}
\end{figure}

For the calculations presented in this section, we adopt an out-of-plane DMI following the CrI$_3$ magnon model~\cite{zhang2026edge}, with $D_z=0.08J_1$. Imposing the degeneracy condition for the two relevant magnon branches gives $J_{2a} = J_{2b} - \frac{2J_c\left(J_1 +3J_{2b}\right)}{3[J_c+6\left(J_1 +3J_{2b}\right)]}$. To characterize the Weyl-magnon regime, we further introduce the inequivalent second-nearest-neighbor exchange couplings, $J_{2a}=0.39$ and $J_{2b}=0.58$. At this value, we project the relevant magnon bands into a $2\times2$ subspace.
The effective Hamiltonian can be expressed as a linear combination of the identity matrix and the Pauli matrices 

\begin{equation}
\begin{split}
H_{eff} = 3\left(J_1 + 3J_{2b}\right) \sigma_0 
+\frac{\sqrt{3}[J_c + 6\left(J_1 + 3J_{2b}\right)]}{4\sqrt{\left(J_1 + 3J_{2b}\right)[J_c +3\left(J_1 + 3J_{2b}\right)]}}
 \left(q_x \sigma_x + q_y \sigma_y\right)
\end{split}
\end{equation}

\begin{figure}[!t]
\centering
\setlength{\figurepanelheight}{0.48\textwidth}
\begin{subfigure}[t]{0.42\textwidth}
\centering
\includegraphics[width=\linewidth]{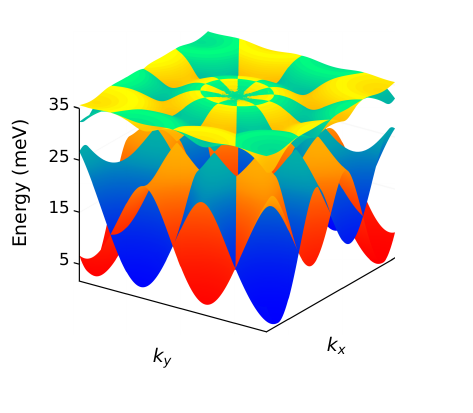}
\caption{}\label{fig:band-a}
\end{subfigure}
\hspace{0.03\textwidth}
\begin{subfigure}[t]{0.42\textwidth}
\centering
\includegraphics[width=\linewidth]{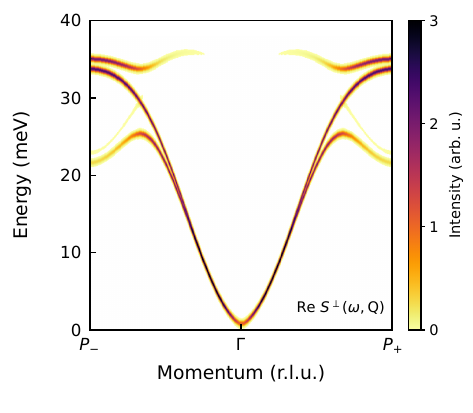}
\caption{}\label{fig:bands-d}
\end{subfigure}
\par\smallskip
\begin{subfigure}[b]{0.42\textwidth}
\centering
\begin{minipage}[c][\figurepanelheight][c]{\linewidth}
\includegraphics[width=\linewidth]{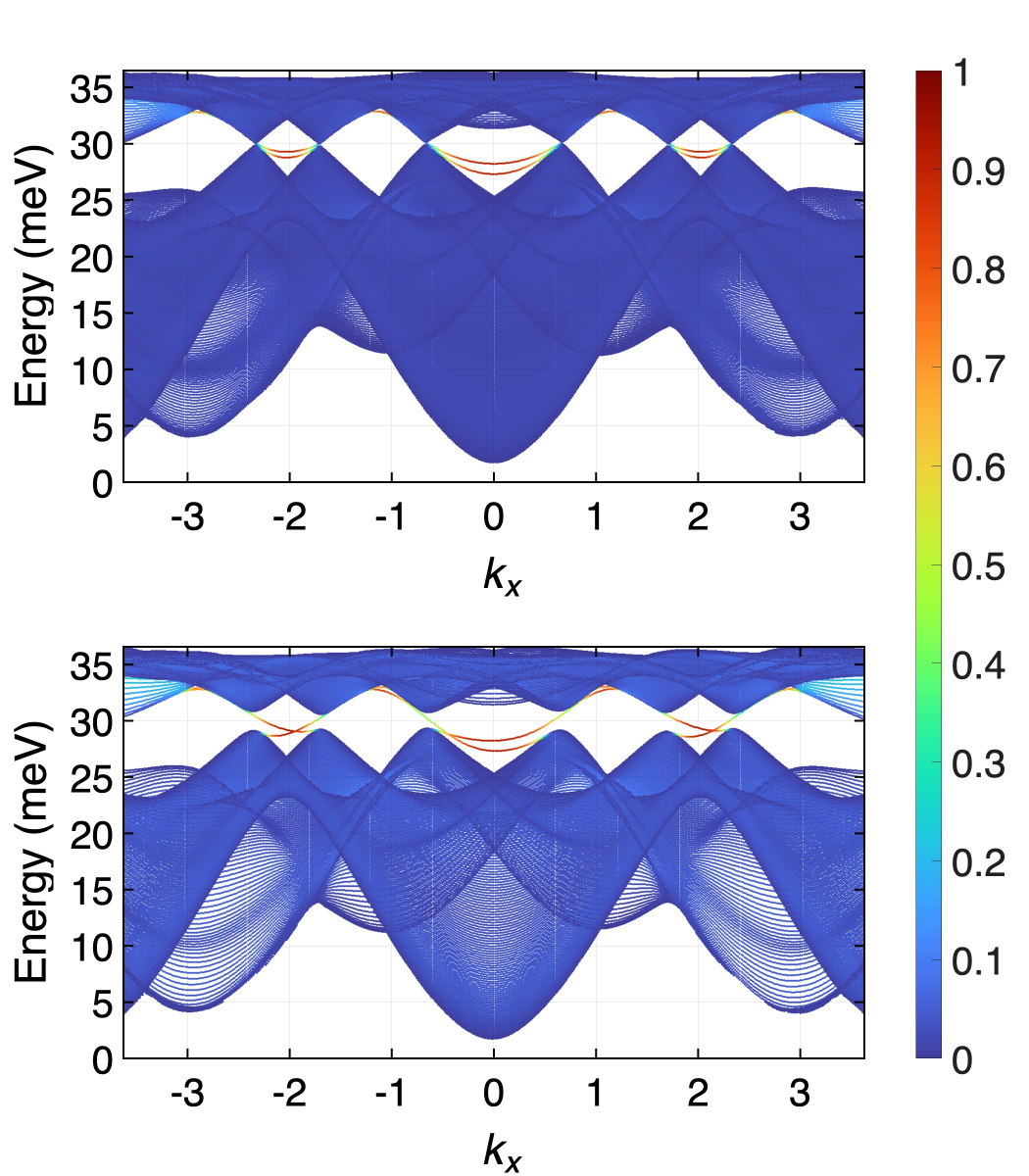}
\end{minipage}
\caption{}\label{fig3c}
\end{subfigure}
\hspace{0.03\textwidth}
\begin{subfigure}[b]{0.42\textwidth}
\centering
\begin{minipage}[c][\figurepanelheight][c]{\linewidth}
\includegraphics[width=\linewidth]{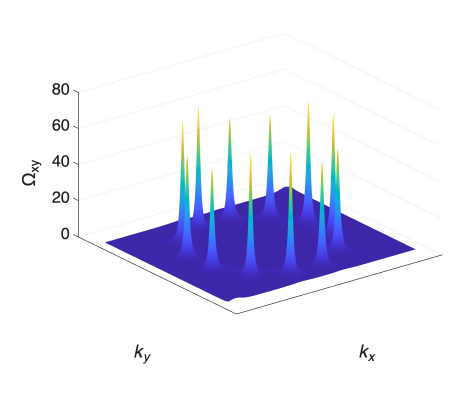}
\end{minipage}
\caption{}\label{fig3d}
\end{subfigure}
\caption{\textbf{Bulk magnon excitations and boundary-localized modes.} (a) Bulk magnon dispersion. (b) Transverse dynamical structure factor, showing the spectral weight of the magnon branches. (c) Slab spectra for two edge configurations, revealing boundary-localized modes between the bulk bands; colour denotes boundary weight. (d) Berry-curvature peaks near the Weyl points, highlighting the geometric response around the band crossings.}
\label{fig3}
\end{figure}

Imposing the BdG biorthonormalization condition $\langle\psi_i|\Xi|\psi_j\rangle = \delta_{ij}$, we obtain the normalized basis:
\begin{align}
|\psi_1\rangle &= \begin{pmatrix} 0 \\ 1 \\ 0 \\ 0 \end{pmatrix},
\qquad |\psi_2\rangle = \frac{1}{\sqrt{4\eta(\eta + J_c)}}\begin{pmatrix} 2\eta + J_c \\ 0 \\ 0 \\ -J_c \end{pmatrix} \equiv \begin{pmatrix} u \\ 0 \\ 0 \\ v \end{pmatrix},
\end{align}

where $\eta=3\left(J_1 + 3J_{2b}\right)$, $u = \frac{2\eta + J_c}{\sqrt{4\eta(\eta + J_c)}}$, and $v = \frac{-J_c}{\sqrt{4\eta(\eta + J_c)}}$, satisfying $u^2 - v^2 = 1$. 
The dynamical spin chirality, $\mathrm{Im}(M^{xy}-M^{yx})$, shown in Fig.~\ref{fig1c} characterizes the circular polarization and chirality of the magnon modes. It is straightforward to evaluate the Dynamic structure factor $S^{\alpha\beta}(\mathbf{k},\omega)$, which is directly related to the inelastic neutron scattering cross section, as shown in Fig.~\ref{fig:bands-d} .
Magnon orbital angular momentum has been studied in collinear magnets and in honeycomb and zigzag lattice models\cite{fishman2022,fishman2023b}.
To analyze the dynamical properties of the excited magnon, we calculate the intrinsic orbital angular momentum carried by the magnons

\begin{equation}
\mathbf{L}_n(\mathbf{k}) = -\frac{\hbar}{2} \sum_{i=1} \left[ |\psi(\mathbf{k})_{i,n}\rangle^{-1} \mathbf{l}_z^{\mathbf{k}} \langle\psi(\mathbf{k})_{i,n}|^{-1} - |\psi '(\mathbf{k})_{i,n}\rangle^{-1} \mathbf{l}_z^{\mathbf{k}} \langle\psi '(\mathbf{k})_{i,n}|^{-1} \right],
\end{equation}

where is $\mathbf{l}_z^{\mathbf{k}}=-i\left(k_x \frac{\partial}{\partial k_y} - k_y \frac{\partial}{\partial k_x}\right)$. 
$|\psi(\mathbf{k})_{i,n}\rangle^{-1}$ denotes the eigenvector component of the $n$-th magnon band at the $i$-th spin site, and $|\psi '(\mathbf{k})_{i,n}\rangle^{-1}$ corresponds to the negative-frequency component. 
The momentum-resolved orbital angular momentum is generally gauge dependent, whereas its angular average provides a gauge-invariant measure\cite{fishman2023}.
Orbital moments can also give rise to a magnon orbital Nernst effect in honeycomb antiferromagnets without SOC \cite{go2024}.
Furthermore, we derive the Berry curvature of the magnon bands from the vector of coefficients multiplying the Pauli matrices $\mathbf{d}(\mathbf{k}) \cdot \boldsymbol{\sigma}$

\begin{equation}
\begin{split}
\rev{\Omega_{xy}^{\pm}}(\mathbf{k}) = \mp \frac{1}{2|\mathbf{d}|^3} \mathbf{d} \cdot \left( \frac{\partial \mathbf{d}}{\partial k_x} \times \frac{\partial \mathbf{d}}{\partial k_y} \right),
\end{split}
\end{equation}

where the superscripts $+$ and $-$ denote the two opposite magnon chiralities, respectively, and the corresponding Chern numbers are obtained by integrating the Berry curvature over the entire first Brillouin zone (BZ), $C_{\pm} = \frac{1}{2\pi} \int_{\mathrm{BZ}} d^2k \, \rev{\Omega_{xy}^{\pm}}(\mathbf{k})$.
The Berry curvature gives magnon wave packets an anomalous transverse velocity, and the spin Nernst response provides a direct transport signature of the unconventional chiral splitting in $i$-wave altermagnetic magnon bands\cite{katsura2010,matsumoto2011,cheng2016,zyuzin2016,cui2023,naka2019}. 
The relevant $C_6\mathcal T$ symmetry relates the two opposite-chirality magnon branches. 
The spin Nernst response stems from the Berry curvature of the magnon band structure. 
The magnon wave packet with a non-zero Berry curvature ($\rev{\Omega_{xy}}$) possesses a self-rotation, leading to an orbital magnetic moment $M_{\Omega_\nu}^z$ that corrects the macroscopic spin current\cite{matsumoto2011prl,matsumoto2011,neumann2020,fishman2022,li2021edh,nie2025edh}.
The intrinsic magnon contribution taking the generic Berry-curvature form

\begin{equation}
{\color{black}\alpha_{xy}^{s}}
=
\frac{k_B}{\hbar}
{\color{black}\sum}
\int_{\mathrm{BZ}}
\frac{d^2k}{(2\pi)^2}
\Xi\,
{\color{black}\Omega_{xy}(\mathbf{k})}\,
c_1\!\left[n_B(\varepsilon_{\mathbf{k}})\right],
\end{equation}

where $c_1$ is the standard thermal weighting function. 
The altermagnetic order is determined jointly by the Néel vector and the crystal symmetry, which constrains the Berry curvature to transform as a sixth-order angular harmonic.
Serving as monopole sources or sinks of the Berry curvature, Weyl magnons give rise to topological edge states and chiral magnon transport against small perturbations\cite{smejkal2022a,smejkal2022b,mook2016,su2017}.

\FloatBarrier

\subsection*{Renormalization and Transport response}

In the presence of magnon--magnon interactions, thermal populations are governed by the eigenstates of the interacting many-body Hamiltonian, whose quasiparticle energies and lifetimes may be renormalized. Quartic terms containing four magnon operators describe magnon--magnon interactions and multimagnon scattering processes. Higher-occupation sectors are neglected when their contributions to the thermal weight and their overlap with the observables of interest are both small. These nonlinear couplings become increasingly important at elevated temperatures, leading to observable effects such as spectral-linewidth broadening and magnon-energy renormalization   
\cite{zhitomirsky2013,mkhitaryan2021,garciagaitan2025}. 
\begin{figure}[!t]
\centering
\setlength{\figurepanelheight}{0.75\textwidth}
\begin{subfigure}[b]{0.43\textwidth}
\centering
\begin{minipage}[c][\figurepanelheight][c]{\linewidth}
\includegraphics[width=\linewidth]{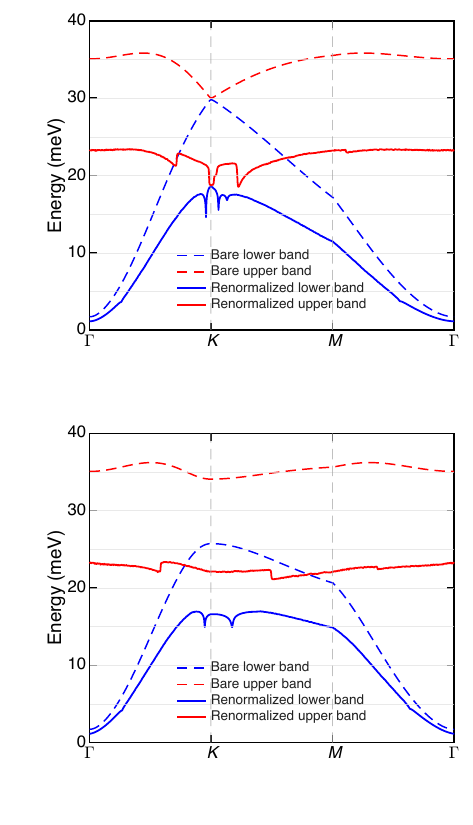}
\end{minipage}
\caption{}\label{fig4a}
\end{subfigure}\hspace{0.03\textwidth}
\begin{subfigure}[b]{0.43\textwidth}
\centering
\begin{minipage}[c][\figurepanelheight][c]{\linewidth}
\includegraphics[width=\linewidth]{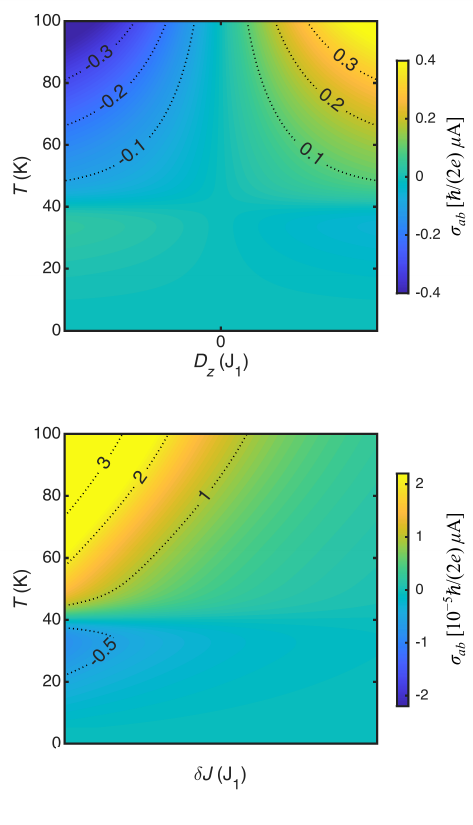}
\end{minipage}
\caption{}\label{fig4b}
\end{subfigure}
\caption{\textbf{Interaction-induced magnon softening and tunable spin transport.} (a) Bare (dashed) and renormalized (solid) dispersions for the two chiralities, showing an overall redshift and momentum-dependent dips. Blue and red denote the lower and upper bands. (b) Extrinsic (top) and intrinsic (bottom) transverse spin conductivities versus temperature and $D_z(J_1)$ or $\delta J(J_1)$, respectively. The extrinsic response changes sign across $D_z=0$, while the intrinsic response exhibits temperature-driven sign reversal.}
\label{fig4}
\end{figure}

Expanding the spin operators to quartic order gives the interaction Hamiltonian

\begin{equation}
\begin{aligned}
-J_{1}\left(a_{m}^{\dagger}a_{m}b_{n}^{\dagger}b_{n}-\frac{a_{m}b_{n}^{\dagger}b_{n}^{\dagger}b_{n}}{4}-\frac{a_{m}^{\dagger}a_{m}b_{n}b_{n}^{\dagger}}{4}-\frac{a_{m}^{\dagger}b_{n}^{\dagger}b_{n}b_{n}}{4}-\frac{a_{m}^{\dagger}a_{m}^{\dagger}a_{m}b_{n}}{4}\right),\\
\end{aligned}
\end{equation}

where the derivation of other coupling parameters is provided in the Supplementary Material. 
Magnon--magnon interactions cause renormalization of the magnon bands, which adjust the energy spectrum in a momentum-dependent manner after Fourier transformation\cite{dyson1956a,zhitomirsky2013,garciagaitan2025}. 
Since the four-field operators cannot generically be solved exactly, the interacting Green's function is determined by contracting pairs of operators with the mean-field averages, where the quartic operators are contracted with the mean-field values of the Bose distribution $n_B(\varepsilon_{\mathbf{q}})=\frac{1}{e^{\varepsilon_{\mathbf{q}}/(k_B T)}-1}$. 
At first order the Hartree contribution is proportional to the original Hamiltonian, yielding a temperature-dependent strength of the mean-field correction. 

We focus on the quartic interaction, which generates the leading nontrivial vertex correction to the magnon propagator. 
For an $\alpha$-magnon carrying momentum $\mathbf{k}$, the leading contribution at the second order contains chirality magnons with a thermal magnon $\mathbf{q}$ from the bath, scattering into two intermediate states with momenta $\mathbf{p}$ and $\rev{\mathbf{k}_r}=\mathbf{k}+\mathbf{q}-\mathbf{p}$.
After transforming into the Bogoliubov basis, the corresponding Matsubara self-energy is


\begin{equation}
\begin{aligned}
\Sigma_2^{\alpha}(\mathbf{k},i\omega_n)=
\left(\frac{1}{N\beta}\right)^2
\sum_{\mathbf{p},\mathbf{q},\omega_u,\omega_v}
\frac{
W(\mathbf{p},\mathbf{q};\mathbf{k})}{
(i\omega_u-\rev{\varepsilon_{\mathbf{p}}})
(i\omega_v-\rev{\varepsilon_{\mathbf{k}_r}})
[i(\omega_u+\omega_v-\rev{\omega_n})-\rev{\varepsilon_{\mathbf{q}}}]
},
\end{aligned}
\end{equation}

where $W(\mathbf{p},\mathbf{q};\mathbf{k})=\left| \Gamma(\mathbf{p},\mathbf{q};\mathbf{k})\right|^2$ is the vertex weight.
After analytic continuation $i\omega_n\rightarrow\omega+i0^+$, we calculate the branch-resolved self-energy for the rainbow approximation. 
The Real part of $\Sigma_2^{\alpha}(\mathbf{k},i\omega_n)$ provides an additional correction to the magnon energy, causing the bands to deform shape. 
The on-shell scattering rate of the $\alpha$-magnon is $-\operatorname{Im}\Sigma_2^{\alpha}(\mathbf{k},i\omega_n)$, which yields a finite lifetime at elevated temperatures. 
Thus, the BdG Hamiltonian determines the bare chiral magnon spectrum and Bogoliubov amplitudes, while the quartic interaction generates the vertex-mediated self-energy, energy shift of the $\alpha$-magnon.
The renormalized magnon bands exhibit pronounced dips at the Weyl magnons, which are quantitatively reproduced only when the second-order self-energy is included. 
The comparison between the bare and renormalized magnon dispersions along a path through the Weyl points (Fig.~\ref{fig4a}) reveals chirality-dependent renormalization: one chirality exhibits band splitting, whereas the other does not.
Despite the substantial interaction-induced band renormalization near the Weyl points, the topological gap remains robust even at high temperatures.

The distinctive feature of the $C_6\mathcal{T}$ altermagnet is that the
chirality-dependent part of the magnon spectrum transforms as a sixth-order
angular harmonic. Since the chirality splitting
$\Delta\varepsilon(\mathbf{k})
=\varepsilon^{+}(\mathbf{k})-\varepsilon^{-}(\mathbf{k})$
is real-valued, its angular dependence can be expanded as
\begin{equation}
\Delta\varepsilon(k,\theta_{\mathbf{k}})
=
\sum_{n}
\left[
a_n(k)\cos(n\theta_{\mathbf{k}})
+
b_n(k)\sin(n\theta_{\mathbf{k}})
\right],
\end{equation}
where $a_n(k),b_n(k)$ are expansion coefficients.
For the $i$-wave state, the leading term is
\begin{equation}
\Delta\varepsilon(\mathbf{k})
=
a_6(k)\cos 6\theta_{\mathbf{k}}
+
b_6(k)\sin 6\theta_{\mathbf{k}}
+\cdots .
\end{equation}

The first-order Hartree correction shifts the magnon bands to lower energies,
whereas the second-order self-energy introduces strongly momentum-dependent
corrections near high-symmetry points. The sixth-order altermagnetic
anisotropy is preserved, while its amplitude and angular phase are
renormalized.

Choosing a crystallographic principal axis as the angular origin, the leading
$i$-wave form factor can be written as
$f_i(\theta_{\mathbf{k}})=\cos 6\theta_{\mathbf{k}}$.
The corresponding spin Nernst response takes the form
\begin{equation}
\alpha_{\mathrm{SN}}^{s}(\theta)
=
A_6\cos 6\theta
+
B_6\sin 6\theta ,
\end{equation}
where $A_6$ and $B_6$ are real coefficients representing the amplitudes of the
$\cos 6\theta$ and $\sin 6\theta$ components of the sixth-order angular harmonic,
respectively.
It therefore satisfies
$\alpha_{\mathrm{SN}}^{s}(\theta+\pi/3)
=\alpha_{\mathrm{SN}}^{s}(\theta)$
and
$\alpha_{\mathrm{SN}}^{s}(\theta+\pi/6)
=-\alpha_{\mathrm{SN}}^{s}(\theta)$.

Finite-temperature transport is renormalized along with the bands themselves. Within linear response, the spin Nernst conductivity $\alpha_{xy}^{s}$ is set by the Berry-curvature-weighted group velocities of the magnon bands, so that a sign change of $\alpha_{xy}^{s}$ reports directly on the thermal reconstruction of the dispersion. Once the self-energy corrections are included, $\alpha_{xy}^{s}$ undergoes two consecutive sign reversals with increasing temperature, a behaviour that we trace to the renormalization of the magnon bands by thermal fluctuations. The reversal coexists with a persistent altermagnetic splitting: magnon--magnon interactions reduce the amplitude of the chiral splitting without altering its sign or its $i$-wave character, and the distinct energies and dispersions of the two chiral branches remain a resource for generating spin-polarised magnon currents\cite{sarkar2025spin}. The same interaction reshapes the spectrum near the Weyl points, where a pronounced dip, reproduced only when the second-order self-energy is included, reduces the local slope of the dispersion and hence the group velocity, which may localize magnon transport in the immediate vicinity of the nodes. Decay sets a second limit on the same excitations. The optical branch is kinematically unstable against decay throughout the Brillouin zone and the acoustic branch becomes unstable in certain regions, so that magnons are formally damped quasiparticles whose finite lifetime curtails the propagation length and degrades spin and energy transport. The decay rate is, however, strongly anisotropic: the restricted phase space keeps the attenuation appreciable only along certain crystallographic directions and negligible along others, and small overall, so that magnons in this regime remain long-lived quasiparticles suitable for transport and device applications.

According to the bulk-edge correspondence, two gapless edge states with opposite group velocities traverse the topological gap, their persistence rooted in the unchanged Chern numbers of the bulk\cite{mook2014,shindou2013,zhang2023tkl,zhang2023sho}. The abrupt termination of the periodic potential adds Tamm--Shockley states whose energy is fixed by an edge chemical potential generated by the anisotropy; this edge potential shifts the dangling edge modes towards the $\Gamma$ point and thereby modifies the edge-mode positions and the magnon density of states. Boundary sublattice termination also controls edge-state energies and bound-state locations in electronic topological systems\cite{kheirkhah2022,zhu2023}, while edge engineering enables spatial relocation and profile tuning of boundary states in an acoustic topological insulator\cite{yang2026}. The Weyl magnons nevertheless remain well defined over a finite temperature range, and a magnetic field combined with the anisotropy aligns the spins more effectively, suppresses the thermal agitation and thereby raises the critical temperature. Because the topological edge states inherit their protection from the bulk topology, they withstand defects and impurities and provide a platform for dissipationless transport of direct relevance to magnon spintronics.


\section*{Discussion}

In bilayer van der Waals (vdW) magnets, specific combinations of stacking offset and interlayer spin alignment selectively break or preserve spin-exchanging symmetry operations. This drives transitions between conventional antiferromagnetic and altermagnetic states, where symmetry alone dictates the existence of chiral-splitting bands and Weyl magnons\cite{liu2022spingroup,chen2024,li2018,yu2026findspingroup,smejkal2020,smejkal2022a,eto2025}. Density functional theory, complemented by symmetry analysis, reveals that the monolayer behaves as a conventional collinear ferromagnet, lacking intrinsic momentum-dependent spin splitting. In this system, rhombohedral and monoclinic stacking patterns couple with the Dzyaloshinskii--Moriya interaction and easy-axis anisotropy to determine the sign of the interlayer exchange, thereby governing the magnetic phase without altering the chemical composition. The branch- and chirality-resolved self-energy complements self-consistent renormalized spin-wave theory and spin-dynamics numerics by isolating the vertex weight $W(\mathbf{p},\mathbf{q};\mathbf{k})$ and so linking the microscopic scattering process to the band deformation. Even where quasiparticles remain well defined and strong spontaneous decay is absent, quantum effects from magnon--magnon interaction that are independent of decay still renormalize the bands, and although the decay rate is small and strongly anisotropic, appreciable only along certain crystallographic directions, the finite lifetime of the optical branch and of parts of the acoustic branch bounds the propagation length and degrades spin and energy transport. The fingerprints are correspondingly accessible, through the presence, symmetry and orientation of the nonrelativistic spin splitting near high-symmetry points, through coupling-induced Raman signatures of the coupled lattice dynamics, and through the linear spin-spin correlation measured by inelastic neutron scattering\cite{chen2018,chen2021,huang2017,huang2018bilayer,schneeloch2024,yu2025vhs,gondolf2025}, with twist angle, strain, gating, layer count and interfacial coupling all acting on this single control\cite{cui2026twisted,ghader2022,zhang2026edge}.

The renormalization that reshapes the bulk bands therefore leaves the topological transport channel intact. For edge states, unchanged bulk Chern numbers leave counterpropagating gapless modes traversing the gap, with the anisotropy-induced edge chemical potential merely repositioning the Tamm--Shockley spectrum and the magnon density of states, while a magnetic field combined with the anisotropy aligns the spins, suppresses thermal agitation, raises the critical temperature and stabilizes the edge modes against decay into the two-magnon continuum\cite{mook2014,shindou2013,zhang2023tkl,zhang2023sho,mook2016,su2017}. The claims are nevertheless bounded. The model is a minimal two-layer anisotropic Heisenberg Hamiltonian parametrized from first principles rather than fitted to spectroscopic data, and the sign of the Nernst response is parameter-sensitive and should be read as a phase-diagram statement rather than a universal value. Given that stacking serves as the control variable, intentional layer arrangement can fold the Brillouin zone with moire superlattices whose long-wavelength modulation bands and locally differing layer registries, could support the interlayer texture acting as the inter-magnon scattering channel examined here. The same symmetry argument should carry to other layered antiferromagnets with tunable interlayer registry, interlayer couplings and spin-orbit strengths to stacks tuned by strain, electrostatic gating. Engineering compensated magnetic states that nonetheless carry a finite spin polarization, is what would make low-stray-field, high-speed and energy-efficient spintronic, memory and quantum-information devices plausible.

\section*{Acknowledgements}
\begingroup\emergencystretch=6em
We thank Jihai Zhang, Dingyong Zhong and  Dele Li for helpful discussions. This work is supported by NKRDPC-2022YFA1402802, NSFC-92165204, the Guangdong Provincial Key Laboratory of Magnetoelectric Physics and Devices (2022B1212010008), the Research Center for Magnetoelectric Physics of Guangdong Province (2024B0303390001), the Guangdong Provincial Quantum Science Strategic Initiative (GDZX2401010 and GDZX2601002), and the Leading Talent Program of Guangdong Special Projects (201626003).\endgroup

\bibliographystyle{cri3-unsrt}
\bibliography{citev4}

\end{document}